\documentclass[graybox]{svmult}

\usepackage{type1cm}        
\usepackage{makeidx}         
\usepackage{graphicx}        
\usepackage{multicol}        
\usepackage[bottom]{footmisc}

\usepackage{newtxtext}       %
\usepackage{newtxmath}       

\usepackage{subfigure}

\newcounter{oldpdfimageresolution}
\newcommand{\includegraphicsdpi}[3]{%
	\setcounter{oldpdfimageresolution}{\pdfimageresolution}
	\pdfimageresolution=#1
	\includegraphics[#2]{#3}%
	\pdfimageresolution=\theoldpdfimageresolution
}

\usepackage{xspace}

\usepackage{units}

\usepackage{color}

\usepackage[numbers,sort&compress]{natbib}

\usepackage[noabbrev, capitalise]{cleveref}
\crefrangeformat{equation}{(#3#1#4) to~(#5#2#6)}
\crefmultiformat{equation}{(#2#1#3)}{ and~(#2#1#3)}{, (#2#1#3)}{ and~(#2#1#3)}

\newcommand{\R}[1]{\mathbb{R}^{#1}}

\newcommand{\C}[1]{\ensuremath{C^{#1}}}
\newcommand{\SO}{SO^3}

\newcommand{\intd}[1]{\mathrm{d}{#1}}

\newcommand{\tns}[1]{\underline{\boldsymbol{#1}}}
\newcommand{\tnss}[1]{\boldsymbol{#1}}
\newcommand{\tnssI}{\tnss{I}}
\newcommand{\tnsO}{\tns{0}}
\newcommand{\tr}{^{\mathrm T}}

\renewcommand{\vv}[1]{\boldsymbol{#1}}

\newcommand{\mat}[1]{\boldsymbol{#1}}
\newcommand{\matO}{\mat{0}}

\newcommand{\br}[1]{\left(#1\right)}

\newcommand{\degree}{\ensuremath{^{\circ}}}

\newcommand{\ie}{i.e.~}

\newcommand{\cf}{cf.~}

\newcommand{\btsvc}{beam-to-solid volume coupling\xspace}

\newcommand{\sr}{Simo--Reissner\xspace}
\newcommand{\kl}{Kirchhoff--Love\xspace}
\newcommand{\eb}{Euler--Bernoulli\xspace}

\newcommand{\cs}[1][]{cross-section#1\xspace}

\newcommand{\ez}{\tns{e}_3}

\newcommand{\osolid}{\Omega^S}
\newcommand{\osolidO}{\osolid_0}

\newcommand{\obeamLO}{\Omega^B_{L,0}}

\newcommand{\beamlength}{L}

\newcommand{\Xsolid}{\tns{X}^S}
\newcommand{\xsolid}{\tns{x}^S}
\newcommand{\usolid}{\tns{u}^S}

\newcommand{\sbeam}{s}

\newcommand{\ubeamr}{\tns{u}^B_r}

\newcommand{\rbeam}{\tns{r}}
\newcommand{\rbeamO}{\tns{r}_0}
\newcommand{\triad}{\tnss{\Lambda}}

\newcommand{\Spk}{\tnss{S}}
\newcommand{\F}{\tnss{F}}
\renewcommand{\E}{\tnss{E}}
\newcommand{\dE}{\delta \E}

\newcommand{\VO}{V_0}
\newcommand{\dV}{\intd{\VO}}

\newcommand{\dWs}{\delta W^S}
\newcommand{\dWb}{\delta W^B}
\newcommand{\dWc}{\delta W^C}

\newcommand{\dWexts}{\delta W^S_{\mathrm{ext}}}

\newcommand{\lagrange}{\tns{\lambda}}

\newcommand{\qsolid}{{\vv{d}^S}}

\newcommand{\qbeam}{\vv{d}^B}

\newcommand{\qlagrange}{\vv{\lambda}}

\newcommand{\resbeam}{\vv{r}^B}
\newcommand{\ressolid}{\vv{r}^S}

\newcommand{\intsolid}[1]{\int_{\osolidO}{ #1 \, \dV}\,}

\newcommand{\intbeamcenterlineO}[1]{\int_{\obeamLO}{ #1 \,\mathrm d \sbeam}\,}

\newcommand{\beamEint}{\Pi^B_{\mathrm{int}}}

\renewcommand{\D}{\mat{D}}

\newcommand{\M}{\mat{M}}

\newcommand{\Kss}{\mat{K}_{SS}}
\newcommand{\Kbb}{\mat{K}_{BB}}

\makeindex             

\begin{document}

\title*{Finite Element Formulations for Beam-to-Solid Interaction -- From Embedded Fibers Towards Contact}
\titlerunning{Finite Element Formulations for Beam-to-Solid Interaction}

\author{Alexander Popp and Ivo Steinbrecher}

\authorrunning{Popp, A., Steinbrecher, I.}

\institute{Popp, A., Steinbrecher, I. \at University of the Bundeswehr Munich, Institute for Mathematics and Computer-Based Simulation (IMCS), Werner-Heisenberg-Weg 39, 85577 Neubiberg, Germany\\ \email{alexander.popp@unibw.de, ivo.steinbrecher@unibw.de}}
%
%
\maketitle
\begin{articlededication}
	Dedicated to Prof. Peter Wriggers on the occasion of his 70th birthday. Since the very first day of his PhD research in computational contact mechanics, the first author has been inspired by the internationally acknowledged, innovative work of Prof. Peter Wriggers in this field (as in many other fields). Later, a successful collaboration has emerged in the organization of scientific events such as the CISM Advanced Course "Computational Contact and Interface Mechanics" in 2016 and the biennial "International Conference on Computational Contact Mechanics (ICCCM)" under the auspices of ECCOMAS.
\end{articlededication}

\abstract*{Contact and related phenomena, such as friction, wear or elastohydrodynamic lubrication, remain as one of the most challenging problem classes in nonlinear solid and structural mechanics. In the context of their computational treatment with finite element methods (FEM) or isogeometric analysis (IGA), the inherent non-smoothness of contact conditions, the design of robust discretization approaches as well as the implementation of efficient solution schemes seem to provide a never ending source of hard nuts to crack. This is particularly true for the case of beam-to-solid interaction with its mixed-dimensional 1D-3D contact models. Therefore, this contribution gives an overview of current steps being taken, starting from state-of-the-art beam-to-beam (1D) and solid-to-solid (3D) contact algorithms, towards a truly general 1D-3D beam-to-solid contact formulation.}

\abstract{Contact and related phenomena, such as friction, wear or elastohydrodynamic lubrication, remain as one of the most challenging problem classes in nonlinear solid and structural mechanics. In the context of their computational treatment with finite element methods (FEM) or isogeometric analysis (IGA), the inherent non-smoothness of contact conditions, the design of robust discretization approaches as well as the implementation of efficient solution schemes seem to provide a never ending source of hard nuts to crack. This is particularly true for the case of beam-to-solid interaction with its mixed-dimensional 1D-3D contact models. Therefore, this contribution gives an overview of current steps being taken, starting from state-of-the-art beam-to-beam (1D) and solid-to-solid (3D) contact algorithms, towards a truly general 1D-3D beam-to-solid contact formulation.}

\section{Introduction}

Computational contact mechanics is a particularly challenging research field in nonlinear solid and structural mechanics that has been shaped over the course of almost four decades now by Professor Peter Wriggers and his many co-workers as well as collaborators \cite{Wriggers2006, Popp2018a}. The interest of most researchers has in large parts been directed towards 3D solid-to-solid contact with the aim of accurately resolving large deformation contact problems of elastic and inelastic engineering components, manufacturing processes or biomechanical systems. Nevertheless, a smaller yet constant stream of innovative contributions on 1D beam-to-beam contact can also be observed, which focus more on the significance of contact interaction of slender rod-wise components such as ropes, cables, fiber webbings or again biological structures on different length scales.

Among the most important research topics for 3D solid-to-solid contact has always been the choice of a robust contact discretization scheme suitable for large deformation kinematics, with some of the famous variants being node-to-segment, Gauss-point-to-segment or segment-to-segment schemes to name only a few \cite{Simo1985,Wriggers1990}. Within the last two decades, so-called mortar methods originally proposed in the context of domain decomposition have emerged as widely accepted state-of-the-art contact discretization approach \cite{Fischer2005, Popp2009, Popp2010, Popp2012a, Popp2014, Farah2015}. Equally important is the question of contact constraint enforcement within the underlying variational formulation, where some common options are penalty methods, Lagrange multiplier methods, the Augmented Lagrangian approach and, more recently, Nitsche's method \cite{Zavarise1999, Wriggers2007}. In recent years, NURBS-based isogeometric analysis has beome increasingly popular also in computational contact mechanics due to its higher-order continuity, which can be advantageous in the accurate resolution of curved contact boundaries \cite{Temizer2011,Lorenzis2014,Seitz2016,Wunderlich2019}. Without any claim of completeness, other relevant research directions in the field of 3D solid-to-solid contact include mesh adaptivity \cite{Wriggers1995}, contact smoothing \cite{Wriggers2001}, third-medium contact \cite{Wriggers2013} and virtual element methods (VEM) \cite{Wriggers2016}.

When considering contact interaction of 1D slender rod-like structures, the associated challenges are quite different, both from a mechanical and from a computational perspective. In particular, the underlying 1D Cosserat continuum theory of beams requires a thorough re-formulation of contact and friction kinematics, which becomes rather complex in the fully nonlinear realm with finite deformations and finite cross-section rotations. Early contributions can be found in \cite{Wriggers1997, Zavarise2000, Litewka2002}. In recent years, such 1D beam-to-beam contact formulation have again been investigated more closely, partly due to progress in the development of geometrically exact beam finite elements and partly due to the high relevance of fiber-based structures and materials for many modern engineering applications \cite{Neto2015,Meier2016paper,Meier2017}.

It is striking that the combined treatment of solid-based contact and beam-based contact in the sense of a 1D-3D beam-to-solid contact formulation has received much less attention over the years, which can of course be attributed to the fact that it combines the complexities associated with both building blocks \cite{Ninic2014,Konyukhov2015,Steinbrecher2017,Humer2020}. The authors of this contribution have recently started research efforts to develop a fully nonlinear 1D-3D beam-to-solid contact formulation including friction. In the following, an overview of the first steps taken in this direction is given, i.e. tied contact formulations for 1D fibers being embedded into 3D solid bodies \cite{Steinbrecher2020}. Moreover, several numerical examples illustrate the interplay of the resulting 1D-3D beam-to-solid volume coupling approach with well-established solid-based contact and beam-based contact models, thus showing the path towards truly general 1D-3D beam-to-solid contact interaction.

\section{Governing Equations}
\label{sec:equations}

\begin{figure}
	\centering
	\includegraphics[scale=1]{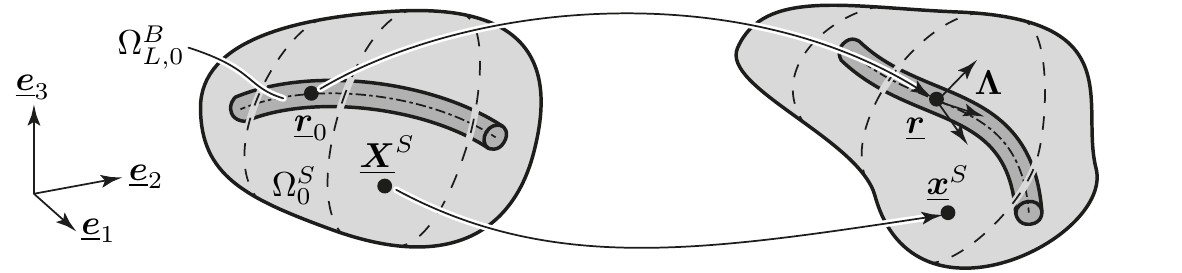}
	\caption{Notation of the finite deformation \btsvc problem.}
	\label{fig:governing_problem}
\end{figure}

We consider a quasi-static 3D finite deformation \btsvc problem, \cf Figure~\ref{fig:governing_problem}.
The weak form is derived by the principle of virtual work with contributions from the solid, the beam, and the coupling terms.

The solid is modeled as a 3D continuum, represented by the open set $\osolidO \subset \R3$ in the reference configuration.
The volume of the embedded beam is not explicitly subtracted from the solid volume, thus resulting in a modeling error due to overlapping material points.
However, the high ratio of beam stiffness to solid stiffness typically alleviates the impact of this error for the envisaged applications.
The deformed position~$\xsolid=\Xsolid + \usolid$ is related to the reference position~$\Xsolid$ through the displacement field~$\usolid$.
Virtual work contributions $\dWs$ of the solid are given by
\begin{equation}
\dWs
=\intsolid{\Spk : \dE}
-\dWexts,
\end{equation}
where~$\delta$ denotes the variation of a quantity, $\Spk$~the second Piola-Kirchhoff stress tensor, $\E$~the energy-conjugate Green-Lagrange strain tensor and $\dWexts$~the virtual work of the external forces.
The Green-Lagrange strain tensor is given as~$\E = \frac{1}{2}(\F\tr \F - \tnssI)$, with $\F = \partial \xsolid/\partial \Xsolid$ being the material deformation gradient and $\tnssI \in \R{3\times 3}$ the identity tensor, respectively.
For the compressible or nearly incompressible solid, we assume a hyperelastic strain energy function~$\Psi(\E)$, which is related to the second Piola-Kirchhoff stress tensor via~$\Spk=\partial \Psi(\E) / \partial \E$.

The beams used in this work are based on the fully nonlinear, geometrically exact beam theory, which in turn builds upon the kinematic assumption of plane, rigid \cs{s}.
The complete beam kinematics can be defined by a centerline curve $\rbeam(\sbeam) \in \R3$, connecting the \cs centroids, and a field of right-handed orthonormal triads $\triad(\sbeam) \in \SO$ defining the rotation of the \cs{s}.
Here $s \in [0, \beamlength] =: \obeamLO \subset \R{}$ is the arc-length along the undeformed beam centerline and $\triad(\sbeam)$ is a rotation tensor, which maps the global Cartesian basis vectors onto the local \cs basis vectors.
The beam centerline displacement $\ubeamr$ relates the undeformed beam centerline position $\rbeamO$ to the deformed position $\rbeam = \rbeamO + \ubeamr$.
The beam contribution to the global virtual work reads
\begin{equation}
\dWb =
\delta \beamEint
-\dWb_{\mathrm{ext}},
\end{equation}
where $\delta \beamEint$ is the variation of the beam's internal elastic energy and $\dWb_{\mathrm{ext}}$ is the virtual work of external forces and moments on the beam.
The elastic beam energy $\beamEint$ depends on the employed beam theory, namely either \sr, \kl or \eb (torsion free) theories, \cf \cite{Meier2019}.

In the real physical problem, the beam surface (defined by the centerline fields $\rbeam$ and $\triad$) is tied to an underlying, internal solid surface.
Modeling this type of surface-to-surface interaction would result in a computationally expensive evaluation of the coupling terms, thus canceling out the advantages of employing a 1D beam theory.
For the inherent assumption in the \btsvc method, that the beam \cs dimensions are small compared to the other dimensions of the problem, we approximate the coupling of the beam surface and the solid volume as a coupling between the beam centerline and the solid volume.
This is a significant change in the mathematical description of the mechanical model, and for the implications of this choice on the applicability and spatial convergence of the proposed method the interested reader is referred to \cite{Steinbrecher2020}.
The kinematic coupling constraints formulated along the beam centerline $\obeamLO$ in the reference configuration read
\begin{equation}
\ubeamr - \usolid = \tnsO \quad \text{on} \quad \obeamLO.
\end{equation}
The constraints are enforced weakly via a Lagrange multiplier field $\lagrange(s) \in \R3$ defined on the beam centerline.
We point out that from a physical point of view $\lagrange$ represents a line load along the beam centerline.
Coupling contributions $\dWc$ to the total virtual work read
\begin{equation}
\dWc =
\intbeamcenterlineO{ \lagrange \br{ \delta \ubeamr - \delta \usolid } }
+
\intbeamcenterlineO{ \delta \lagrange \br{ \ubeamr - \usolid } }.
\end{equation}
This leads to the final saddle point-type weak formulation of the 1D-3D \btsvc problem,
\begin{equation}
\dWs + \dWb + \dWc = 0.
\end{equation}

\section{Spatial Discretization}
\label{sec:discretization}

We build upon the finite element method here for spatial discretization.
In the solid field either standard $\C0$-continuous finite elements or a $\C1$-continuous isogeometric approach based on second-order NURBS is used.
The beam centerlines are discretized using $\C1$-continuous finite elements based on third-order Hermite polynomials.
For details on the objective and path-independent interpolation of the rotational field along the beam centerline, see \cite{Meier2019}.
Employing a mortar-type coupling approach \cite{Wohlmuth2000}, the Lagrange multipliers are also approximated with a finite element interpolation.
In this work, linear Lagrange polynomials are used to interpolate the Lagrange multiplier field, \cf \cite{Puso2004}, thus yielding discrete mortar coupling matrices $\D$ and $\M$ associated with the beam and solid side, respectively.
In the spirit and nomenclature of classical mortar methods, the beam is treated as the slave side and the solid as the master side, respectively.
The linearized system of equations to be solved in every Newton iteration exhibits saddle-point structure and reads
\begin{equation}
\label{eq:discret_global_system}
\begin{bmatrix}
\Kss & \matO & -\M\tr \\
\matO & \Kbb & \D\tr \\
-\M & \D & \matO
\end{bmatrix}
\begin{bmatrix}\Delta \qsolid \\ \Delta \qbeam \\ \qlagrange
\end{bmatrix}
=
\begin{bmatrix}
-\ressolid \\
-\resbeam \\
\M \qsolid -\D \qbeam
\end{bmatrix},
\end{equation}
where $\qsolid$, $\qbeam$, $\Delta \qsolid$ and $\Delta \qbeam$ are the displacements and their increments of solid and beam, $\ressolid$ and $\resbeam$ denote the discrete residual vectors, $\Kss=\partial \ressolid /  \partial \qsolid$ and $\Kbb=\partial \resbeam /  \partial \qbeam$ are the tangent stiffness matrices, and $\qlagrange$ are the discrete Lagrange multiplier values.
The system \eqref{eq:discret_global_system} is solved by introducing a weighted penalty regularization.

\section{Numerical Examples}
\label{sec:examples}

\begin{figure*}[tbp]
	\centering
	\subfigure[Reference configuration.]{\includegraphicsdpi{300}{}{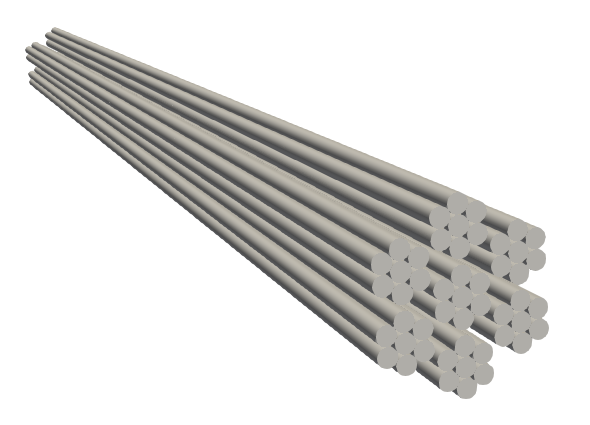}\label{fig:ex_rope_ref}}
	\hfil
	\subfigure[Configuration at load step 40.]{\includegraphicsdpi{300}{}{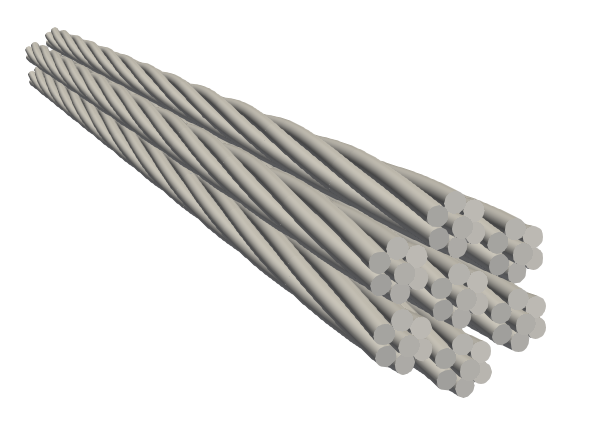}\label{fig:ex_rope_stage_1}}	
	\subfigure[Configuration at load step 60.]{\includegraphicsdpi{300}{}{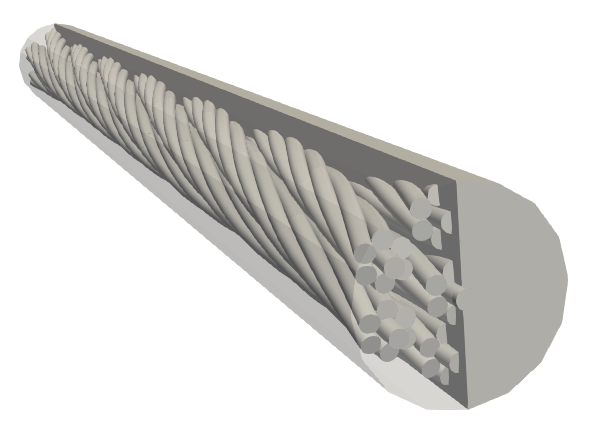}\label{fig:ex_rope_stage_2}}
	\hfil
	\subfigure[Configuration at load step 80.]{\includegraphicsdpi{300}{}{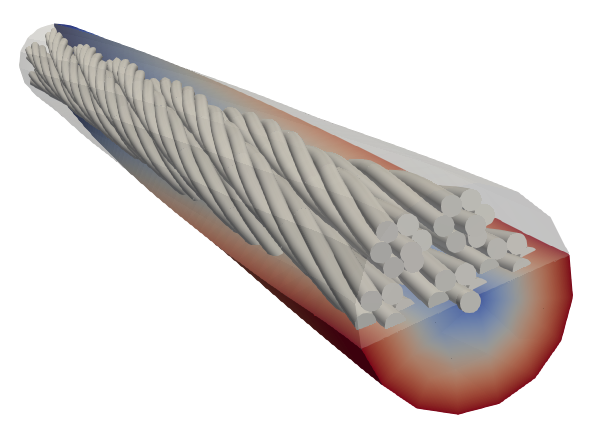}\label{fig:ex_rope_stage_3}}
	\caption{
		Configuration of the coated rope example at different stages during the simulation.
	}
	\label{fig:ex_rope}
\end{figure*}
\begin{figure*}[tbp]
	\sidecaption
	\centering
	\includegraphics[scale=1]{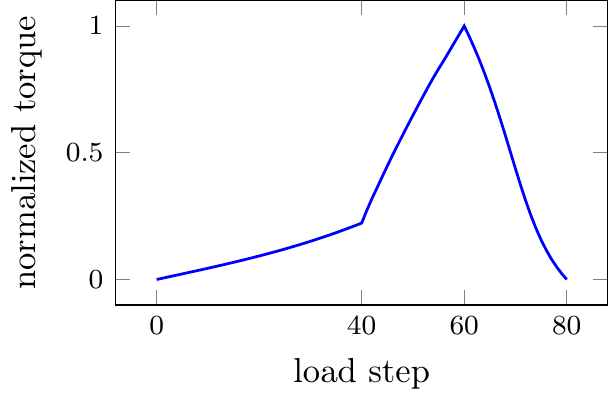}
	\caption{
		Resulting (normalized) torque at the fixed end of the rope over the course of the simulation.
	}
	\label{fig:ex_rope_torque}
\end{figure*}

First, we investigate the twisting of a coated rope consisting of $7 \times 7$ individual fibers.
This is an extension of the rope example proposed in \cite{Meier2017} and, unless stated otherwise, all parameters are taken from said reference.

The initial configuration of the 49 straight fibers is shown in Figure~\ref{fig:ex_rope_ref}.
Each fiber has the length $L=\unit[2.5]{m}$ and is discretized using 10 torsion free beam elements.
The All-Angle Beam Contact (ABC) formulation \cite{Meier2017} is used to model the beam-to-beam contact which arises in this example.
In the first stage of the simulation the fibers are loaded in axial direction and each of the seven sub-bundles of fibers is twisted around its center fiber by two full rotations.
This twisting process is realized in a Dirichlet controlled manner.
The solution is obtained within 40 quasi-static load steps and is displayed in Figure~\ref{fig:ex_rope_stage_1}.
In the next stage the sub-bundles themselves are rotated around the rope axis by an additional full rotation, \cf Figure~\ref{fig:ex_rope_stage_2}.
Again this is realized with Dirichlet boundary conditions and the solution is obtained within 20 additional quasi-static load steps.
Up to this point the model consists only of beam elements.
In the next and final stage the twisted wires are covered with a solid coating ($E_{\text{solid}} = E_{\text{beam}} / 200$).
The coating and the (deformed) beams are coupled to each other via the \btsvc method using first-order interpolation of the Lagrange multipliers.
The Dirichlet conditions at the end of the fibers are replaced by Neumann boundaries matching the reaction forces at the Dirichlet conditions in the last step of the twisting process.
Therefore, the bundle of wires is in equilibrium with itself and the external loads, \ie there is no initial interaction between the solid coating and the wires.
The Neumann load at the end of the fibers is now linearly reduced to zero within 20 quasi-static load steps.
Now the solid coating interacts with the beam fibers in the sense of a pre-stressed composite material.
This results in a back-twisting of the rope and coating of about a quarter rotation as displayed in Figure~\ref{fig:ex_rope_stage_3}.
The normalized (to a maximum value of 1) reaction torque at the fixed end of the wire is plotted in Figure~\ref{fig:ex_rope_torque}.
One can observe that even tough the external loads are decreased linearly (in load steps 60--80) the resulting reaction torque exhibits a nonlinear behavior, thus highlighting the complexity of this problem.
This example illustrates the combination of beam-to-beam contact with \btsvc and also shows the maturity of the \btsvc method for complex beam and solid geometries as well as real life engineering problems.
Therefore, it can be seen as an important step towards a truly general 1D-3D beam-to-solid contact formulation.

\begin{figure*}[tbp]
	\centering
	\subfigure[Reference configuration.]{\includegraphics[scale=1]{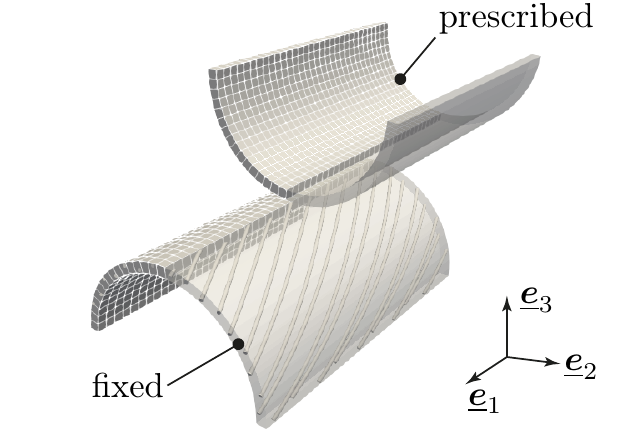}\label{fig:ex_curved_composite_000}}
	\hfil
	\subfigure[Configuration at load step 100.]{\includegraphicsdpi{300}{}{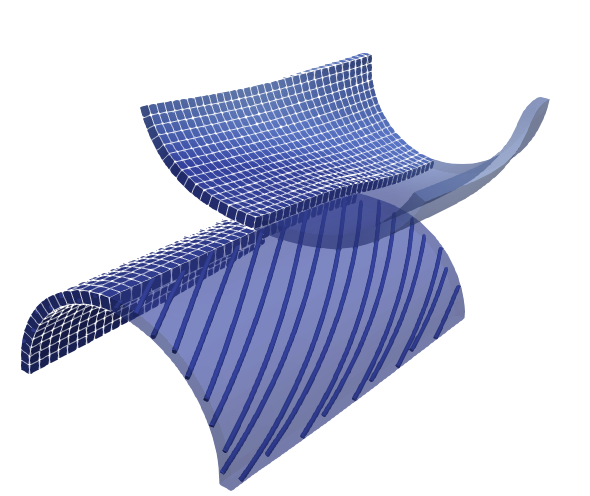}\label{fig:ex_curved_composite_100}}	
	\subfigure[Configuration at load step 150.]{\includegraphicsdpi{300}{}{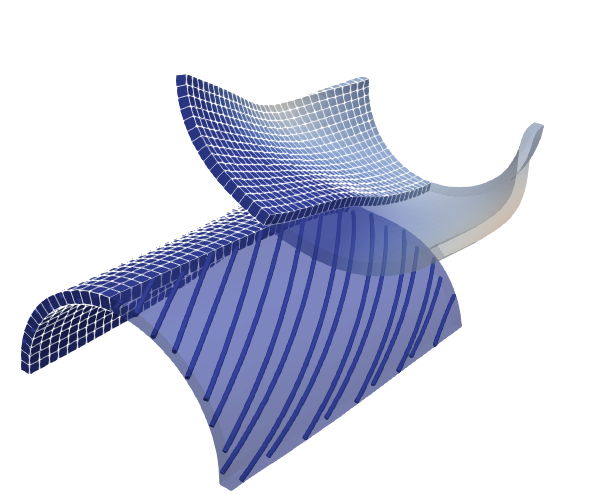}\label{fig:ex_curved_composite_150}}
	\hfil
	\subfigure[Configuration at load step 200.]{\includegraphicsdpi{300}{}{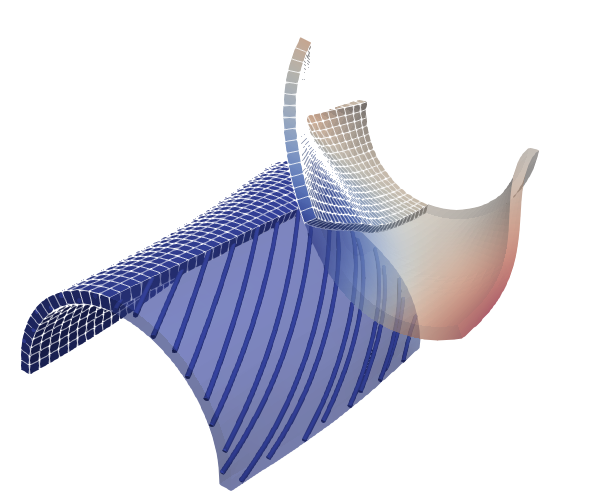}\label{fig:ex_curved_composite_200}}
	\caption{
		Configuration of the reinforced shell example during different stages of the simulation.
	}
	\label{fig:ex_curved_composite}
\end{figure*}
In the second example, two shells with the shape of hollow half-cylinders are pressed against each other.
The two shells have the same spatial dimensions as well as material parameters, but one of them is reinforced with stiff ($E_{\text{fiber}} = 200 E_{\text{shell}}$) fibers, which are oriented at $45\degree$ relative to the cylinder axis.
The reference configuration is shown in Figure~\ref{fig:ex_curved_composite_000}.
The reinforced shell is fixed at one end and the other shell is moved towards the fixed one with prescribed displacements in negative~$\ez$ direction.
Solid-shell finite elements \cite{Bischoff1997, Vu-Quoc2003a} are used to discretize the shells, and both shells employ the same hyperelastic Saint Venant--Kirchhoff material.
The reinforcements are explicitly discretized with beam finite elements and coupled to the shell elements via the \btsvc method.
The 3D solid-to-solid (surface-to-surface) contact between the shells is modeled with a state-of-the-art mortar approach using standard Lagrange multipliers \cite{Popp2010,Popp2014}.
The problem is solved using 200 quasi-static load steps and \cref{fig:ex_curved_composite_100,fig:ex_curved_composite_200} show the configuration of the two shells at different stages during the simulation.
The reinforcement effects can clearly be seen as the reinforced shell deforms less than the other one.
Moreover, the effect of the asymmetry introduced by the reinforcements can be observed, as the initial geometric symmetry of the problem is broken and the top shell slides down one side of the reinforced shell.
This example showcases the applicability of the \btsvc method to model fiber-reinforced composites and also gives a glimpse on the possible applications in combination with solid-to-solid contact.
Again, it can therefore be seen as an important step towards a truly general 1D-3D beam-to-solid contact formulation.

\section{Conclusion}
Within this contribution we have presented state-of-the-art finite element formulations for beam-to-solid interaction.
Specifically, slender fibers are modeled using efficient 1D beam theories and subsequently embedded inside 3D solid volumes with a mortar-type coupling approach.
This allows to efficiently combine different (\ie 1D and 3D) modeling approaches into a mixed-dimensional finite element formulation, which has been demonstrated with two illustrative examples.
The presented framework is by no means limited to embedded fibers, but is currently extended towards beam-to-solid contact problems as part of our ongoing research.

\bibliographystyle{spmpsci}      
\bibliography{literature,literature_steinbrecher,literature_popp}   

\end{document}